\documentclass{appolb}
\usepackage{epsfig,wrapfig,amsmath,amssymb,cite}

 \def\rat{r}                    
 \def\rl{\varkappa}             

\begin{document}
\title{Axial masses in quasielastic neutrino scattering and
       single-pion neutrinoproduction on nucleons and nuclei.%
\thanks{Presented at the XXth Max Born Symposium ``Nuclear effects in neutrino
        interactions,'' Wroc{\l}aw, Poland, December 7--10, 2005, by V.A.~Naumov.
       }
      }
\author{Konstantin S.~Kuzmin%
        \thanks{Also Institute for Theoretical and Experimental Physics, 
         RU-117259 Moscow, Russia.},
        Vladimir V.~Lyubushkin%
        \thanks{Also Physics Department of Irkutsk State University,
             RU-664003 Irkutsk, Russia.}, \\
         Vadim A.~Naumov
        \thanks{Also INFN, Sezione di Firenze, I-50019 Sesto Fiorentino, Italy.}
\address{Joint Institute for Nuclear Research, RU-141980 Dubna, Russia}
       }
\maketitle

\begin{abstract}
We analyse available experimental data on the total charged-current ${\nu}N$
and $\overline{\nu}N$ cross sections for quasielastic scattering and single-pion
neutrinoproduction.
Published results from the relevant experiments at ANL, BNL, FNAL, CERN,
and IHEP are included dating from the end of sixties to the present day,
covering $\nu_{\mu}$ and $\overline{\nu}_{\mu}$ beams on a variety of nuclear
targets, with energies from the thresholds to about $350$~GeV.
The data are used to adjust the poorly known values of the axial masses.
\end{abstract}


\section{Introduction}

It is well known that the theoretical description of the cross sections for
CC and NC (quasi)elastic neutrino-nucleon scattering (QES) and single-pion
neutrinoproduction through baryon resonances (RES) are very sensitive to
the shape of the weak axial-vector elastic and transition form factors.
By adopting the standard dipole parametrization for these form factors,
their shapes can be described with the two phenomenological parameters
$M_A^{\text{QES}}$ and $M_A^{\text{RES}}$, the so-called axial (dipole) masses.
In general, these masses are different and, moreover, the numerical value of
$M_A^{\text{RES}}$ is vastly dependent of the particular dynamic model for
the resonance production.

The experimental values for both $M_A^{\text{QES}}$ and $M_A^{\text{RES}}$
coming from measurements of (quasi)elastic neutrino and antineutrino scattering
off protons and nuclei and from the more involved and model-dependent analyses
of charged pion electroproduction off protons, spread within rather wide ranges.
In this study we attempt to fine-tune the axial masses by fitting all
available data on the CC QES (with $\Delta Y=0$) and RES $1\pi$
total cross sections for ${\nu}_\mu$ and $\overline{\nu}_\mu$ scattering off
different nuclear targets from experiments at
ANL  \cite{Kustom:69,Campbell:73,Mann:73,S.Barish:77,Radecky:82},
BNL  \cite{Baker:81,Kitagaki:86},
FNAL \cite{Bell:78,Barish:80,Kitagaki:83,Asratyan:84,Suwonjandee:04}
CERN \cite{Burmeister:65,Franzinetti:66,Young:67,Eichten:73,Bonetti:77,%
           Krenz:78,Lerche:78,Armenise:79-Pohl:79,Bolognese:79,Allen:80,%
           Allasia:83,Allen:86,Jones:89,Allasia:90}, and
IHEP \cite{Makeev:81,Belikov:82,Belikov:85,Grabosch:88,Grabosch:89,Brunner:89,%
           Ammosov:92}.
In our opinion, this procedure is more selfconsistent in comparison with the usual
straightforward averaging over the experimental values of the axial mass
(see, e.g., Ref.~\cite{Bernard:01}) extracted under different assumptions about
the other badly known factors involved into the analyses of each experiment.
 
\section{Axial mass from the data on quasielastic scattering}
\label{QES}

\begin{figure}[t!]
\centering
\epsfig{file=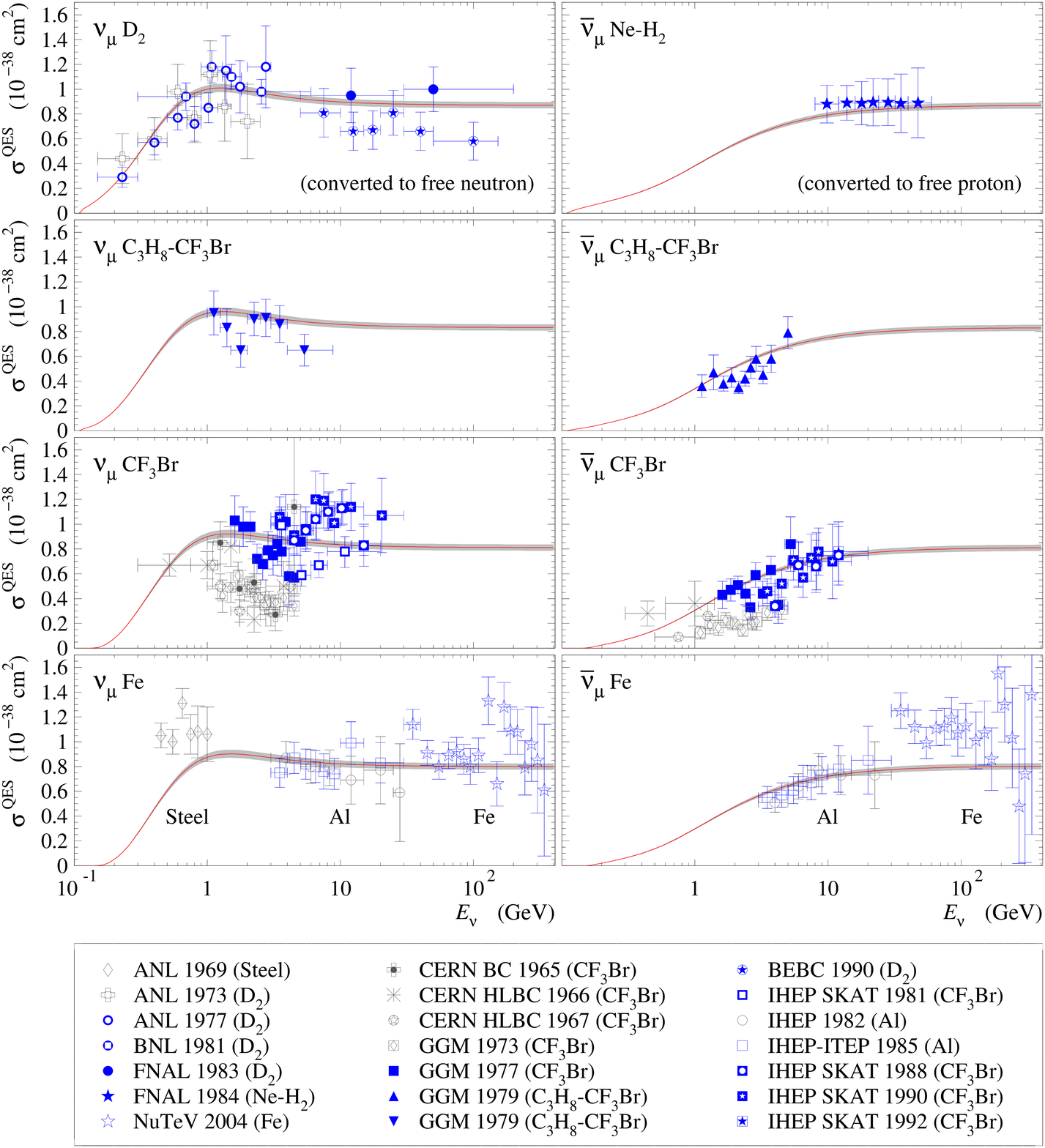,width=\linewidth}
\caption{Total quasielastic $\nu_{\mu}n$ and $\overline{\nu}_{\mu}p$ cross sections
         measured for different nuclear targets by the experiments
         ANL 1969 \cite{Kustom:69},
         ANL 1973 \cite{Mann:73},
         ANL 1977 \cite{S.Barish:77},
         BNL 1981 \cite{Baker:81},
         FNAL 1983 \cite{Kitagaki:83},
         FNAL 1984 \cite{Asratyan:84},
         NuTeV 2004 \cite{Suwonjandee:04},
         CERN BC 1965 \cite{Burmeister:65},
         CERN HLBC 1966 \cite{Franzinetti:66},
         CERN HLBC 1967 \cite{Young:67},
         GGM 1973 (Gargamelle, CERN) \cite{Eichten:73},
         GGM 1977 \cite{Bonetti:77},
         GGM 1979 \cite{Armenise:79-Pohl:79},
         BEBC 1990 (CERN) \cite{Allasia:90},
         IHEP SKAT 1981 \cite{Makeev:81},
         IHEP 1982 \cite{Belikov:82},
         IHEP-ITEP 1985 \cite{Belikov:85},
         IHEP SKAT 1988 \cite{Grabosch:88},
         IHEP SKAT 1990 \cite{Brunner:89}, and
         IHEP SKAT 1992 \cite{Ammosov:92}.
         The curves and bands correspond to the world average value of
         $M_A^{\text{QES}}=0.95\pm0.03~\text{GeV}$ obtained with the GKex(02S)
         model for the vector form factors from the fit to the subset of
         these data (160 datapoints).
         The data of Refs.~\cite{Kustom:69,Mann:73,Burmeister:65,%
         Franzinetti:66,Young:67,Eichten:73,Belikov:82} (39 grey datapoints) are
         rejected from the fit being either superseded or not satisfying our 
         selection criterion $\chi^2/\text{ndf}<4.5$.
        }
\label{Fig:Sigma_tot_QES}
\end{figure}

Figure~\ref{Fig:Sigma_tot_QES} shows a compilation of the QES data from experiments at
ANL  \cite{Kustom:69,Mann:73,S.Barish:77},
BNL  \cite{Baker:81},
FNAL \cite{Kitagaki:83,Asratyan:84,Suwonjandee:04},
CERN \cite{Burmeister:65,Franzinetti:66,Young:67,Eichten:73,Bonetti:77,%
           Armenise:79-Pohl:79,Allasia:90}, and
IHEP \cite{Makeev:81,Belikov:82,Belikov:85,Grabosch:88,Brunner:89,Ammosov:92}
performed with a variety of nuclear targets.
The cross sections reported in the earlier experiments
\cite{Kustom:69,Mann:73,Burmeister:65,Franzinetti:66,Young:67,Eichten:73}
exhibit uncontrollable systematic errors and fall well outside the most probable
range determined through the fit to the full dataset of about 200 datapoints; the
value of $\chi^2$ evaluated for each subset of these data exceeds $\sim5~\text{ndf}$.
Hence, following the (nonstringent) selection criterion $\chi^2/\text{ndf}<4.5$,
they were excluded from the final fit.

For the ${\nu}n\to\mu^-p$ and $\overline{\nu}p\to\mu^+n$ cross sections we use
the result of Ref.~\cite{Kuzmin:04a} neglecting possible second-class
current contributions (see Appendix); under this standard assumption it coincides
with that of Strumia and Vissani~\cite{Strumia:03}.
For the elastic electromagnetic form factors we apply the QCD VM model of Gari and
Kr\"uempelmann~\cite{Gari:92} extended and fine-tuned by Lomon~\cite{Lomon:02}
(``GKex(02S)'' version) and the most current inverse polynomial parametrization by Budd
\emph{et~al.}~\cite{Bradford:06}  (``BBBA2006'') obtained through a global fit to the
world data on the Sachs form factors.
For the axial and pseudoscalar form factors we use the conventional
representations~\cite{LlewellynSmith:72}
\begin{equation}\label{F_A}
F_A\left(Q^2\right)=F_A(0)\left(1+\frac{Q^2}{M^2_A}\right)^{-2},
\quad
F_P\left(Q^2\right)=\frac{2M_N^2}{m^2_\pi+Q^2}F_A\left(Q^2\right),
\end{equation}
with $F_A(0)=g_A=-1.2695\pm0.0029$~\cite{Eidelman:04} (assuming $g_V=1$) and
$M_A \equiv M_A^{\text{QES}}$ being a free parameter of our fit.

The nuclear effects for the data obtained for deuterium
\cite{Mann:73,S.Barish:77,Baker:81,Kitagaki:83,Allasia:90}
and neon-hydrogen \cite{Asratyan:84} targets were subtracted by the authors
of the experiments. Therefore these data are fitted by the cross sections
evaluated for free nucleons. To describe the remaining experimental data we apply
the relativistic Fermi gas model by Smith and Moniz \cite{Smith:72} with
the kinematics and values of binding energies and Fermi momenta of the target
proton and neutron determined by the composition of each target quoted
in Fig.~\ref{Fig:Sigma_tot_QES}.

\clearpage 

\clearpage 

The resulting world average obtained are
\begin{align}
M_A^{\text{QES}}[\text{GKex(02S)}]&=0.95 \pm 0.03~\text{GeV}
\quad(\chi^2/\text{ndf}=0.92), 
\label{M_A_best_fit_GK} \\
M_A^{\text{QES}}[\text{BBBA2006}]&=0.96 \pm 0.03~\text{GeV}
\quad(\chi^2/\text{ndf}=0.91).
\label{M_A_best_fit_BBBA}
\end{align}
The errors correspond to the usual one-standard-deviation errors (MINUIT default
\cite{James:94}) plus the systematic errors, added quadratically, which account for
the uncertainties in the data on the vector form factors, nuclear effects (within
the adopted model) and radiative corrections.
The fit performed, for a comparison, with the naive dipole model for the vector
form factors yields $M_A^{\text{QES}}=0.93 \pm 0.03~\text{GeV}$ with
$\chi^2/\text{ndf}=0.95$.

\begin{wrapfigure}{r}{0.475\linewidth}
\hfill 
\epsfig{file=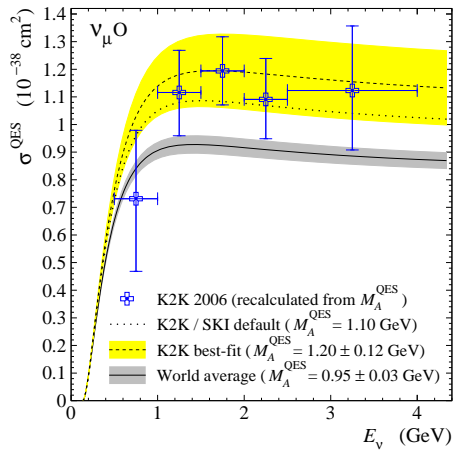,width=0.99\linewidth}
\caption{Comparison between the QES $\nu_{\mu}n_{\text{O}}\to\mu^-p$ cross sections
         (where $n_{\text{O}}$ is a neutron bound in oxygen) evaluated with the best fit
         values \protect\eqref{M_A_best_fit_GK} and \protect\eqref{M_A_best_fit_K2K},
         the current K2K/SK\,I default, and the K2K result reconstructed from the
         fit values of $M_A^{\text{QES}}$ for five energy bins reported in Fig.~9 of
         Ref.~\cite{Gran:06}.
        }
\label{Fig:Sigma_tot_QES_K2K}
\end{wrapfigure}
The obtained world average values of the axial mass are in strong contradiction
with the recently published result of the K2K Collaboration \cite{Gran:06}:
\begin{equation}\label{M_A_best_fit_K2K}
M_A^{\text{QES}}[\text{K2K}]=1.20\pm0.12~\text{GeV}
\end{equation}
This value has been determined for a water target through fitting the $Q^2$ distributions
of muon tracks reconstructed from neutrino-oxygen quasielastic interactions by
using the combined K2K-I and K2K-IIa data from the Scintillating Fiber detector in the
KEK accelerator to Kamioka muon neutrino beam.%
\footnote{Data from the continuation of the K2K-II period were not
          used in the analysis~\cite{Gran:06}.
          The best-fit values of $M_A^{\text{QES}}$ obtained from the K2K-I and
          K2K-IIa data subsets separately are, respectively,
          $1.12\pm0.12$ and
          $1.25\pm0.18~\text{GeV}$.
         }
In Fig.~\ref{Fig:Sigma_tot_QES_K2K} we show the $\nu_{\mu}n\to\mu^-p$ cross section
recalculated from the fit values of $M_A^{\text{QES}}$ obtained in Ref.~\cite{Gran:06}
for the shape of the $Q^2$ distribution for each reconstructed neutrino energy.%
\footnote{The authors underline that the result for each energy should not be
          considered a measurement, but rather a consistency test.}
The calculation was performed with our default inputs that introduces an
uncertainty of at most $2\%$ which is added to the quoted error bars quadratically.
Also shown are the cross sections evaluated by using the world average
value \eqref{M_A_best_fit_GK}, the K2K best fit \eqref{M_A_best_fit_K2K}, 
and the value of $1.1$~GeV used as a default in the recent neutrino oscillation
analyses of K2K~\cite{Ahn:04,Ahn:06} and Super-Kamiokande I \cite{Ashie:05}.

\section{Axial mass from the data on single pion neutrinoproduction}
\label{RES}

\begin{figure}[b]
\centering
\epsfig{file=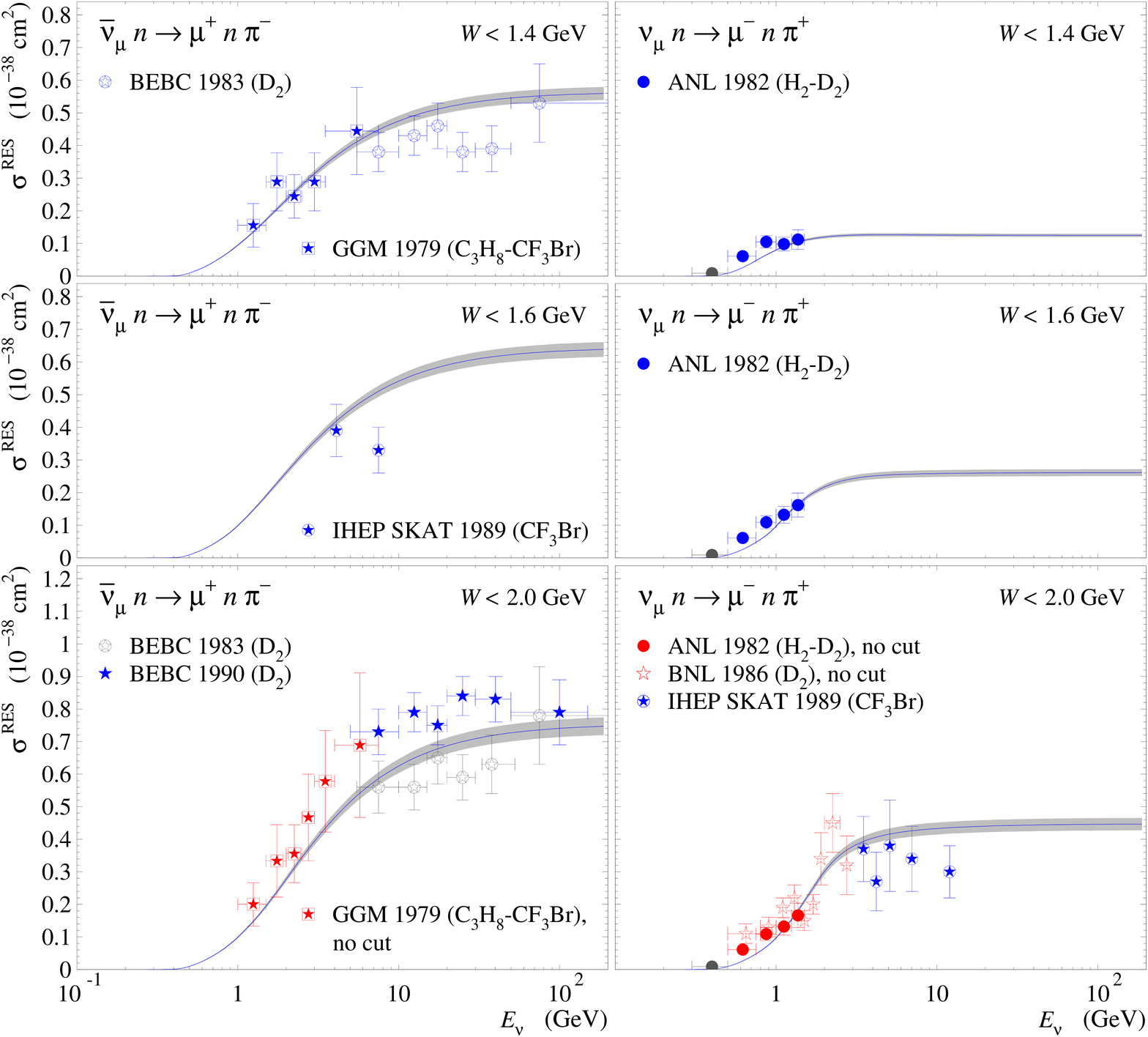,width=\linewidth}
\caption{Total $\pi^-$ and $\pi^+$ production cross sections
         measured for different nuclear targets by the experiments
         ANL 1982 \cite{Radecky:82},
         BNL 1986 \cite{Kitagaki:86},
         GGM 1979 \cite{Bolognese:79},
         BEBC 1983 \cite{Allasia:83},
         BEBC 1990 \cite{Allasia:90}, and
         IHEP SKAT 1989 \cite{Grabosch:89}.
         The data are classified according to the cuts in $W$.
         The curves and bands correspond to the world average value of
         $M_A^{\text{RES}}=1.12\pm0.03~\text{GeV}$ obtained from the fit
         to a subset (196 points) of the full data presented
         in this and two next figures (see text for more details).
        }
\label{Fig:Sigma_tot_RES1}
\end{figure}

\begin{figure}[htb]
\centering
\epsfig{file=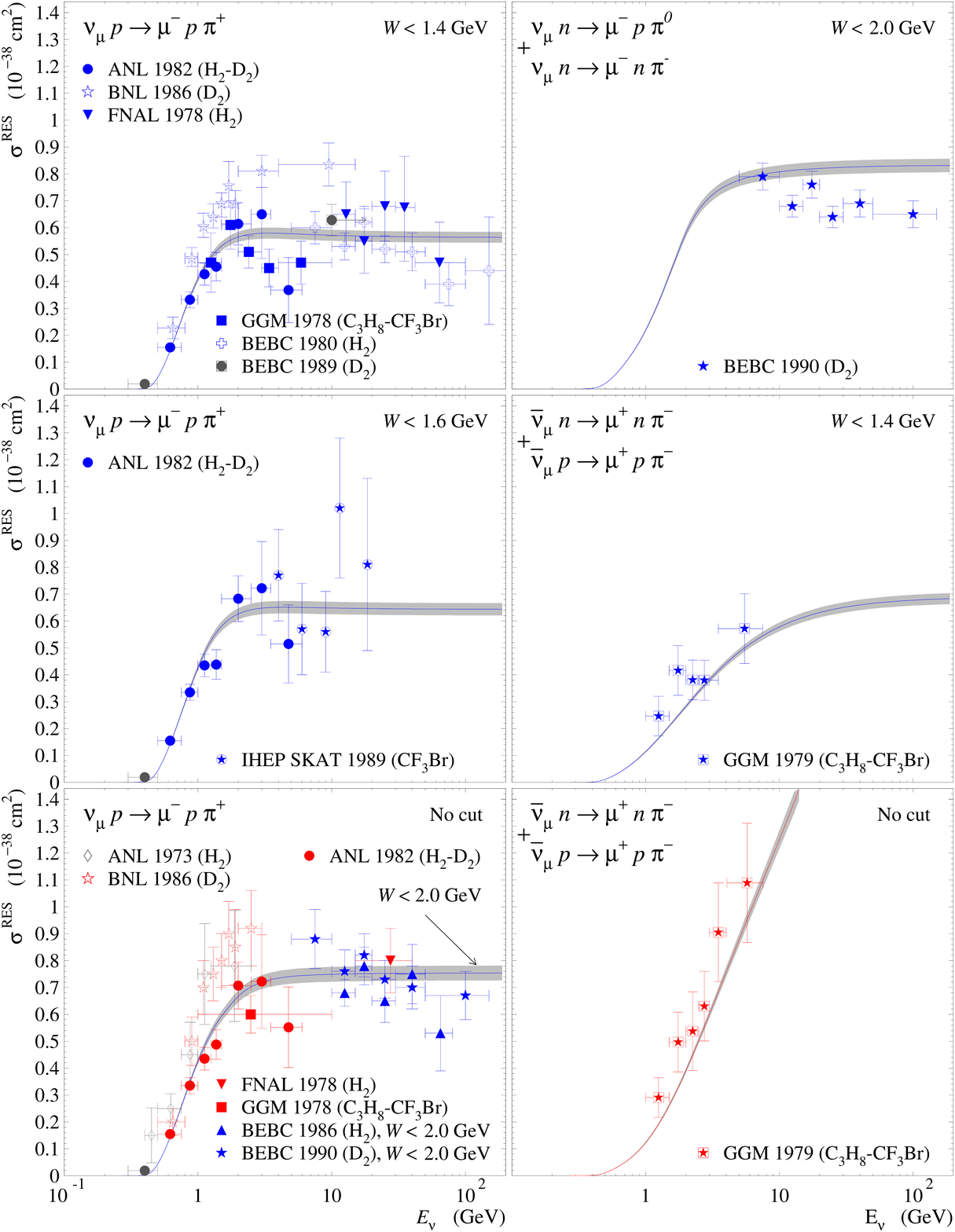,width=\linewidth}
\caption{Total $\pi^-$, $\pi^0$, and  $\pi^+$ production cross sections
         measured for different targets by the experiments
         ANL 1973 \cite{Campbell:73},
         ANL 1982 \cite{Radecky:82},
         BNL 1986 \cite{Kitagaki:86},
         FNAL 1978 \cite{Bell:78},
         GGM 1978 \cite{Lerche:78},
         GGM 1979 \cite{Bolognese:79},
         BEBC 1980 \cite{Allen:80},
         BEBC 1986 \cite{Allen:86},
         BEBC 1989 \cite{Jones:89},
         BEBC 1990 \cite{Allasia:90}, and
         IHEP~SKAT~1989~\cite{Grabosch:89}.
         See Fig. \protect\ref{Fig:Sigma_tot_RES1} and text.
        }
\label{Fig:Sigma_tot_RES2}
\end{figure}

\begin{figure}[htb]
\centering
\epsfig{file=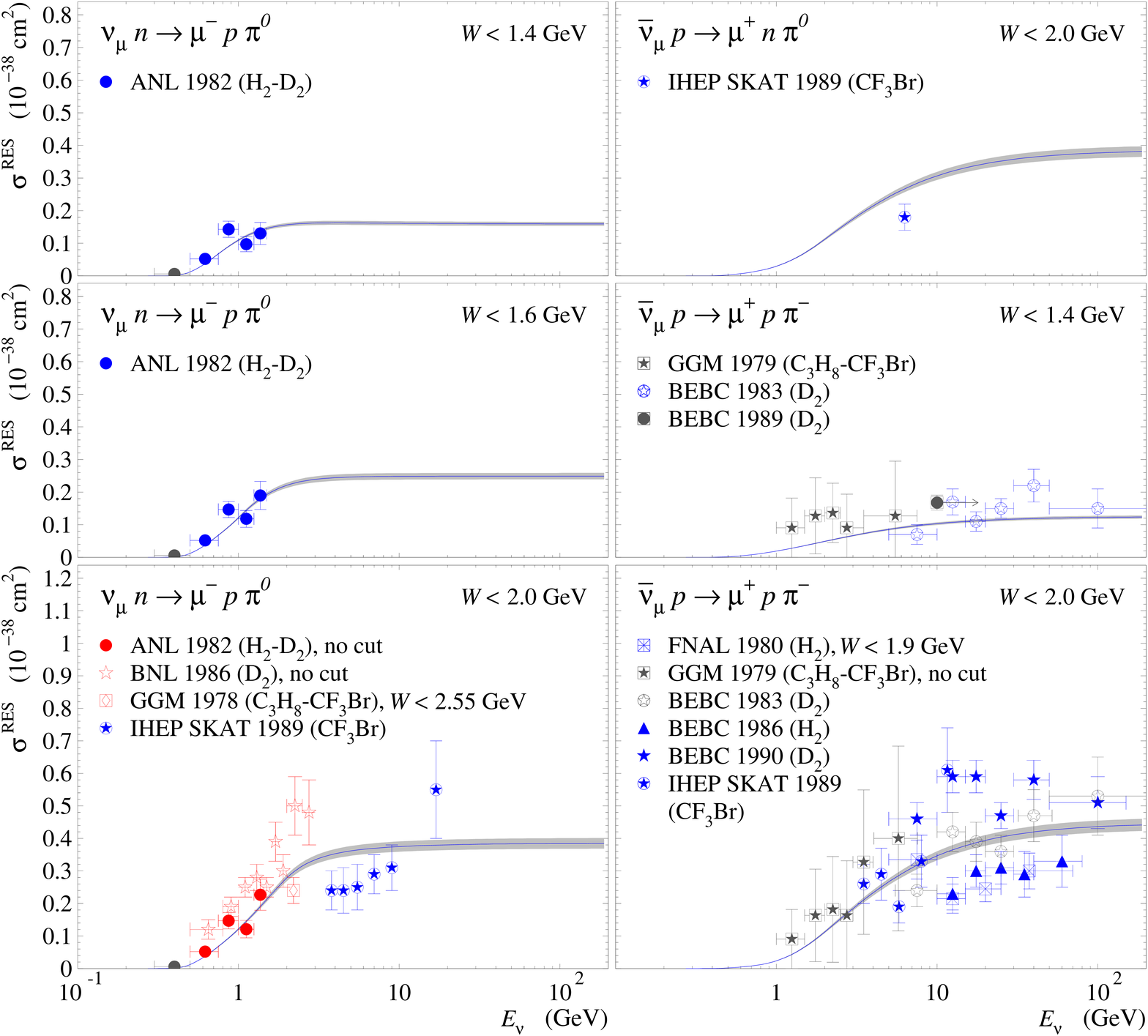,width=\linewidth}
\caption{Total $\pi^0$ and  $\pi^-$ production cross sections
         measured for different  targets by the experiments
         ANL 1982 \cite{Radecky:82},
         BNL 1986 \cite{Kitagaki:86},
         FNAL 1980 \cite{Barish:80},
         GGM 1978 \cite{Krenz:78},
         GGM 1979 \cite{Bolognese:79},
         BEBC 1983 \cite{Allasia:83},
         BEBC 1986 \cite{Allen:86},
         BEBC 1989 \cite{Jones:89},
         BEBC 1990 \cite{Allasia:90}, and
         IHEP SKAT 1989 \cite{Grabosch:89}.
         See Fig. \protect\ref{Fig:Sigma_tot_RES1} and text.
        }
\label{Fig:Sigma_tot_RES3}
\end{figure}

Figures~\ref{Fig:Sigma_tot_RES1}-\ref{Fig:Sigma_tot_RES3} show a compilation
of the data on single pion neutrinoproduction cross sections from experiments at
ANL \cite{Campbell:73,Radecky:82},
BNL \cite{Kitagaki:86},
FNAL \cite{Bell:78,Barish:80},
CERN \cite{Krenz:78,Lerche:78,Bolognese:79,Allen:80,Allasia:83,Allen:86,%
           Jones:89,Allasia:90}, and
IHEP \cite{Grabosch:89}. The nuclear targets are listed in the legends. All the
data, as well as the theoretical curves, are classified through the panels corresponding
to the experimental cut-offs in invariant hadronic mass $W$ ranging from $1.4$ to
$2.55$~GeV and including the measurements without cuts in $W$.

\clearpage 

\clearpage 

For the theoretical description of the single-pion neutrinoproduction through baryon
resonances we apply an extended version of the Rein-Sehgal (RS) model~\cite{Rein:81}.
Our extension~\cite{Kuzmin:RS} is based upon a covariant form of the charged leptonic
current with definite lepton helicity and takes into account the lepton mass.
In the present calculations, we use the same set of 18th nucleon resonances with central
masses below $2$~GeV and the same ansatz for the nonresonance background as in the original
RS model. With that, all relevant parameters are updated according to the current
data~\cite{Eidelman:04}.
Significant factors (normalization coefficients etc.) estimated in Ref.~\cite{Rein:81}
numerically are recalculated by using the new data and a more accurate integration
algorithm.
The relativistic quark model of Feynman, Kislinger, and Ravndal~\cite{FKR}
adopted in the RS approach unambiguously determines the structure of the transition
amplitudes involved into the calculation and the only unknown structures are the
vector and axial-vector transition form factors $G^{V,A}\left(Q^2\right)$.
In the RS model, they are assumed to have the form 
\begin{equation}\label{G_VA}
G^{V,A}\left(Q^2\right)\propto\left(1+\frac{Q^2}{4M_N^2}\right)^{1/2-n}
\left(1+\frac{Q^2}{M_{V,A}^2}\right)^{-2}
\end{equation}
with the ``standard'' value of the vector mass $M_V=0.84~\text{GeV}$ (that is the same
as in the dipole parametrization of the elastic vector form factor).
The integer $n$ in the first (``ad hoc'') factor of Eq.~\eqref{G_VA}
is the number of oscillator quanta present in the final resonance.
The axial mass $M_A=M_A^{\text{RES}}$ (fixed to be $0.95~\text{GeV}$
in the RS model) is the free parameter of our fit.
In order to compensate for the difference between the experimental value of the nucleon
axial-vector coupling $g_A$ and the $SU_6$ predicted value ($g_A(SU_6)=-5/3$),
Rein and Sehgal introduced a renormalization factor $Z=0.75$.
For adjusting the renormalization to the current world averaged value
$g_A=-1.2695\pm0.0029$~\cite{Eidelman:04} we use $Z=0.762$ and assume $g_V=1$.

The nuclear effects for all nuclear targets different from hydrogen and deuterium
are taken into account through the standard Pauli blocking factor (see, e.g.,
Ref.~\cite{Paschos:03} and references therein). The estimated relevant uncertainty
is taken into account in the fit and in the error of its output.

Almost all the data (196 points) shown in Figs.
\ref{Fig:Sigma_tot_RES1}-\ref{Fig:Sigma_tot_RES3}
participate in the fit. Several data subsets are excluded since they are superseded
in the posterior reports of the same collaborations
(e.g., the data from Refs.~\cite{Campbell:73,Allasia:83}),
or are transformation of the others derived from the same experimental samples
(e.g., the data of Refs.~\cite{Bolognese:79} with no cut on $W$).
Note that all the data included into the fit satisfy the criterion $\chi^2/\text{ndf}<4.5$.
The resulting world average obtained in the fit is
\begin{equation}\label{M_A_RES_best_fit}
M_A^{\text{RES}}=1.12 \pm 0.03~\text{GeV}
\quad(\chi^2/\text{ndf}=1.14). 
\end{equation}
As in the QES case, the error is the combination of the $1\sigma$ deviation given by MINUIT
and estimated systematic uncertainties. The obtained world average is in agreement with
the recent analysis by Furuno \emph{et~al.} \cite{Furuno-Sakuda:03} of the BNL 7-foot bubble
chamber deuterium data%
\footnote{The analysis of Ref.~\cite{Furuno-Sakuda:03} is based on the total event sample
          of 1.8~M pictures and holds two periods of runs in 1976-77 and 1979-80.
          The outputs of the analysis are
          $M_A^{\text{RES}}=1.08 \pm 0.07~\text{GeV}$ (statistical error only) -- from the
          fit of the $Q^2$ distributions of $p\pi^+n_s$ events and
          $M_A^{\text{RES}}=1.15_{-0.06}^{+0.08}~\text{GeV}$ (both statistical and QES
          errors are included) -- from the $1\pi$ and QES cross sections ratio.
          The best-fit value of $M_A^{\text{QES}}$ obtained in the same analysis assuming
          the dipole model for the vector form factors (with the standard $M_V$) is
          $1.07 \pm 0.05~\text{GeV}$ that is well above our result (see Sect.~\ref{QES}).
         }
as well as with the value $M_A^{\text{RES}}=1.1~\text{GeV}$ (the same as $M_A^{\text{QES}}$)
adopted in the most recent K2K neutrino oscillation analysis \cite{Ahn:06} but considerably
lower than the value $M_A^{\text{RES}}=1.2~\text{GeV}$ used for the atmospheric neutrino
analysis of the Super-Kamiokande Collaboration \cite{Ashie:05}.
 
\begin{figure}[htb]
\centering
\epsfig{file=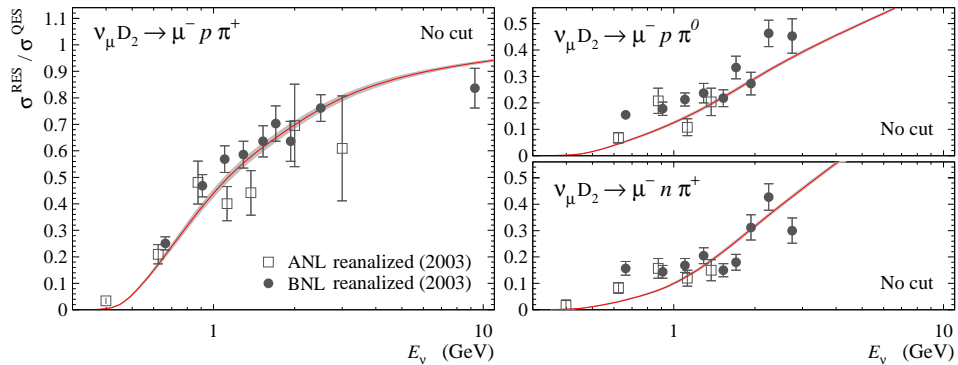,width=\linewidth}
\caption{The ratios of $1\pi$ and QES cross sections evaluated in Ref.~\cite{Furuno-Sakuda:03}
         from the data of ANL and BNL deuterium experiments. The curves and bands are
         calculated with the world average values  \protect\eqref{M_A_best_fit_GK} and
         \protect\eqref{M_A_RES_best_fit} for $M_A^{\text{QES}}$ and $M_A^{\text{RES}}$,
         respectively.
        }
\label{Fig:Ratios}
\end{figure}

Figure \ref{Fig:Ratios} shows a comparison between our calculations and the result of
Ref.~\cite{Furuno-Sakuda:03} for the ANL and BNL data on the ratios of the one-nucleon
normalized $1\pi$ and QES $\nu_\mu\text{D}_2$ cross sections (calculated and measured with
no cut on $W$). Being transformations of the others, these data are not included into
the fit. The narrow bands indicate the uncertainties in the values of the axial masses
\eqref{M_A_best_fit_GK} and \eqref{M_A_RES_best_fit}. The agreement is reasonably good.


\section{Conclusions}

To summarise, we performed a statistical study of the QES and $1\pi$ neutrinoproduction
total cross section data in order to extract the best-fit values of the parameters
$M_A^{\text{QES}}$ and $M_A^{\text{RES}}$. Our results given by Eqs.~\eqref{M_A_best_fit_GK},
\eqref{M_A_best_fit_BBBA}, and \eqref{M_A_RES_best_fit} are, of course, model dependent
and can be recommended for use only within the same (or numerically equivalent) model
assumptions as in the present analysis. We are planning to extend the analysis by employing
more sophisticated treatments of the nuclear effects and including additional experimental
information.

\section*{Acknowledgments}
 
We would like to acknowledge many useful conversations with Krzysztof Graczyk, Jan Sobczyk,
and Oleg Teryaev.

\clearpage 

\section*{Appendix}

The most general formula for the QES ${\nu}N$ cross section is 
\begin{equation*}
\frac{d\sigma^{\text{QES}}}{dQ^2}
=\frac{G^2_F\cos^2\theta_CM^2}{2\pi E^2_\nu}
\left(1+\frac{Q^2}{M_W^2}\right)^{-2}
\left[A+\left(\frac{s-u}{4M^2}\right)B
       +\left(\frac{s-u}{4M^2}\right)^2C\right],
\end{equation*}
where $s=\left(k+p\right)^2$, $u=\left(k'-p\right)^2$, $Q^2=-q^2$;
$k$, $k'=k-q$, and $p$ are the 4-momenta of (anti)neutrino, final lepton,
and initial nucleon, respectively; the coefficient functions $A$, $B$,
and $C$ are given by
\begin{align*}
A = &\  2\left[\left(x'+\rat^2\right)\left(2x'+\rl^2\right)-\rl^4\right]
         \text{Re}\left(F_V^*F_M\right)                                  \\
    &\ -4\rl^2\left\{\rat\text{Re}\left[F_A^*\left(F_V+F_M\right)\right]
       + \left(x'+\rat^2+\rl^2\right)\text{Re}
                                           \left(F_A^*F_P\right)\right\} \\
    &\ + \left[\left(x'+\rl^2\right)\left(x'-1+\rat^2-\rl^2\right)
       - \rat^2\right]\left|F_V\right|^2                                 \\
    &\ + \left[\left(x'+\rl^2\right)\left(x'+1-\rat^2-\rl^2\right)
       - \rat^2\right]\left|F_A\right|^2                                 \\
    &\ - \left[x'\left(x'+\rat^2\right)\left(x'-1+\rl^2\right)
       + \rl^4\right]\left|F_M\right|^2                                  \\
    &\ +4\rl^2\left(x'+\rl^2\right)\left(x'+\rat^2\right)
         \left|F_P^2\right|                                              \\
    &\ \pm4\rat\left(x'+\rat^2\right)\left[\left(x'+1+\rl^2\right)
         \text{Re}\left(F_T^*F_A\right)
       +2\rl^2\text{Re}\left(F_T^*F_P\right)\right]                      \\
    &\ \pm4\rat\rl^2\left[\left(x'+1+\rl^2\right)
         \text{Re}\left(F_S^*F_V\right)
       + \rl^2\text{Re}\left(F_S^*F_M\right)\right]                      \\
    &\ -4\left(x'+\rat^2\right)\left[\left(x'+\rl^2\right)
         \left(x'+1+\rat^2\right)+\rat^2\right]\left|F_T\right|^2        \\
    &\ +4\rl^2\left(x'+1\right)\left(x'+\rl^2\right)\left|F_S\right|^2,  \\
B = &\   \mp4x'\text{Re}\left[F_A^*\left(F_V+F_M\right)\right]
       \pm2\rat\rl^2\left[\left|F_M\right|^2
       + \text{Re}\left(F_V^*F_M+2F_A^*F_P\right)\right]                 \\
    &\ +4\rl^2\text{Re}\left\{F_T^*
         \left[F_A-2\left(x'+\rat^2\right)F_P\right]
       - F_S^*\left(F_V-x'F_M\right)\right\},                            \\
C = &\  \left|F_V\right|^2+\left|F_A\right|^2+x'\left|F_M\right|^2
       \mp4\rat\text{Re}\left(F_T^*F_A\right)
       +4\left(x'+\rat^2\right)\left|F_T\right|^2,
\end{align*}
with the upper (lower) signs corresponding to neutrino
(antineutrino) scattering. The six form factors $F_i$ involved
are functions of $Q^2$;
\[
x=\frac{Q^2}{2(pq)},
\quad
x'=\frac{Q^2}{4M^2},
\quad
\rl=\frac{m}{2M},
\quad
\rat=\frac{M_n-M_p}{2M},
\quad
M=\frac{M_p+M_n}{2},
\]
and the remaining notation is standard. In the limit $M_n=M_p$, the general
formula reduces to that of Ref.~\cite{LlewellynSmith:72} and by putting $F_S=F_T=0$
it coincides with the result of Ref.~\cite{Strumia:03} derived for the inverse
$\beta$ decay, taking account the proton-neutron mass difference.%
\footnote{In fact, the latter effect is insignificant for the present analysis,
          although it is included for completeness, together with the exact
          kinematics.}
In this paper, we apply the Standard Model assumptions ($T$ and $C$
invariance + CVC), thus neglecting the scalar and tensor form factors $F_{S,T}$
induced by the second-class currents, as well as the imaginary parts of the
first-class form factors $F_{V,M,A,P}$.

\clearpage 

\end{document}